\documentclass[twocolumn,showpacs,amsmath,amssymb,pre]{revtex4}
\usepackage{graphicx}% Include figure files

\begin{document}
%\draft
\title{
Anomalous heat conduction and anomalous diffusion in nonlinear lattices, single walled nanotubes, and billiard gas channels}
\author{
Baowen Li$^a$ and Jiao Wang$^b$, Lei Wang$^b$, and Gang Zhang$^a$}
\affiliation{
$^a$Department of Physics, National University of Singapore,
117542 Singapore \\
$^b$ Temasek Laboratories, National University of Singapore, 119260 Singapore
}
\date{13 October 2004}

\begin{abstract}
We study anomalous heat conduction and anomalous diffusion in low dimensional systems ranging from nonlinear lattices, single walled carbon nanotubes, to billiard gas channels.
We find that in all discussed systems, the anomalous heat conductivity can be connected with the anomalous diffusion, namely, 
if energy diffusion is
$\sigma^2(t)\equiv \langle\Delta x^2\rangle =2Dt^{\alpha} (0<\alpha\le 2)$,  then
the thermal conductivity can be expressed in terms of the system
size $L$ as $\kappa = cL^{\beta}$ with $\beta=2-2/\alpha$. This
result predicts that a normal diffusion ($\alpha =1$) implies a
normal heat conduction obeying the Fourier law ($\beta=0$), a
superdiffusion ($\alpha>1$) implies an anomalous heat conduction
with a divergent thermal conductivity ($\beta>0$), and more
interestingly, a subdiffusion ($\alpha <1$) implies an anomalous
heat conduction with a convergent thermal conductivity
($\beta<0$), consequently, the system is a thermal insulator in
the thermodynamic limit. Existing numerical data support our
theoretical prediction.
\end{abstract}
\maketitle
%%%%%%%%%%%%%%%%%%%%%%%
{\bf 
``{\it It seems there is no problem in modern physics for which there are on record as many false starts, and as many theories which overlook some essential feature, as in the problem of the thermal conductivity of nonconducting crystals}"\cite{Peierls}. This statement is still true even if half century has past.  In fact, many questions and many puzzles remain open, for example, what is a necessary and sufficient condition for heat conduction in low dimensional system to obey the Fourier law is still not clear.
Nevertheless, some encouraging progress have been achieved through computational simulations and  analytical studies in recent years. We now understand that in a momentum conserved one dimensional system, such as the Fermi-Pasta-Ulam model, the thermal conductivity diverges with the system size, $L$, as $L^{2/5}$, while it is $L^{1/3}$ if the  coupling with the transverse mode is present. A connection between the anomalous heat conduction and anomalous diffusion has been found. Some interesting models for control heat flow such as thermal rectifier, thermal diode, and thermal transistor have been proposed and studied.
}

\section{Introduction}

Heat conduction in low dimensional systems has been attracting increasing attentions in the last decade (see Review
\cite{Review} and the references therein.) The problem is of great interests not only from fundamental point of view but also from application point of view. On the one hand, it is still an open question that under what dynamical condition(s) will the heat conduction obey the Fourier law. It was suggested that the exponential instability might be the necessary condition\cite{Casati84}, and this conjecture has been further confirmed by the study of heat conduction in a quasi one dimensional Lorenz gas channel\cite{Alonso99}. However, the studies on disordered Ehrenfest gas channel\cite{LWH02}, triangle gas channel\cite{LCW03}, polygonal gas channel\cite{Alonso02}, and an alternate mass hard-core potential chain\cite{LCWP04} show that a system with zero Lyapunov exponent can also display a normal diffusion and the heat conduction obeys the Fourier law.

Unlike the billiard gas channel, we still don't have a clear answer to this question in lattice models.
It has long been believed that the nonintegrability might be a sufficient condition for the Fourier law. However, the study on the Fermi-Pasta-Ulam (FPU) model\cite{FPUKm,FPULepri,FHLZ98,HLZ00,Aoki01} shows that the heat conductivity diverges with system size $L$ as $L^{\beta}$ ($\beta>0$), which means that the heat conduction violates the Fourier law. It is clear now that the anomalous thermal conductivity is due to the solitary wave propagation along the chain\cite{FHLZ98}. It has been shown theoretically from the mode-coupling theory\cite{FPUMode} and the Peierls equation \cite{FPUPer} that the divergent exponent $\beta$ should be $2/5$, which is supported by the numerical results from different groups\cite{FPULepri,HLZ00}. For a generic one dimensional lattice, we now understand from the conclusion given by Prosen and Campbell\cite{Prosen00} that the momentum conservation will lead to a divergent thermal conductivity. 

The $2/5$ power law divergence for one dimensional nonlinear lattice looks like quite general. However, an analysis from the renormalization group argue that the divergent exponent should be $1/3$ for a momentum conserved one dimensional system\cite{NR02}. This looks like a puzzle at first glance. However, if we look carefully at the analysis\cite{NR02}, we find that, a random noise added to each particle of the lattice is not possible in a one dimensional system. Recently, Wang and Li\cite{WL04} clarified this puzzle via mode-coupling theory by studying a quasi 1d systems by allowing the lattice have both longitudinal and transverse motions. According to Wang and Li's theory, the $1/3$ power law comes from the mode coupling between the longitudinal modes and transverse modes. When the coupling is absent, like in the 1D FPU model, the $2/5$ is recovered.

If the momentum conservation is broken by introducing an on-site potential like the Frenkel-Kontorova (FK) model\cite{HLZ98} and the $\phi^4$ model\cite{HLZ00,Aoki00}, the heat conduction then obeys the Fourier law.

On the other hand, with the fast development of nanotechnology, many low dimensional systems such as the nanotubes, nanowires and nanobelt etc can be fabricated in the laboratory. It is necessary to know the properties such as the mechanical properties, electronic properties, and thermal properties before they can be put into any practical use. However, compared with other two properties, much less is known about the thermal ones\cite{Nano}. In fact, the very basic question such as whether the heat conduction in such systems can obey the Fourier law is still not clear, even though an increasing studies have been devoted to\cite{Nanoheat,Maruyama,YWLL04,ZL04}. 

Moreover, the understanding of the underlying heat conduction mechanism will shed lights on the application for designing novel thermal devices. Indeed, some promising works have been done in this direction. The first model for thermal rectifier was proposed by Terrano, Peyrard, and Casati\cite{rectifier}. This model allows heat flows anisotropically along the chain. Most recently, based on two coupled FK chains, a thermal diode model is proposed\cite{diode}. This model acts like a conductor when a temperature gradient is applied, while it acts like an insulator when the gradient is reversed. In addition, in a certain parameter range, a negative differential thermal resistance is observed. Based on this property, a thermal transistor\cite{transistor} model is introduced to control the heat flow by adjusting either temperature at or current through the third terminal, just like we do for charge flow in a MOSFET (Metal-Oxide-Semiconductor Field Effect Transistor).  

In this paper, we give a detailed discussion on the anomalous heat conduction and anomalous diffusion in different systems: nonlinear lattices, single walled nanotubes, and billiard gas channels. The main analytical result, Eq. (\ref{alpha-beta}), for the relation between anomalous heat conduction and anomalous diffusion has been reported in Ref\cite{LW03}. The paper is organized as the follows. In Sec II, we give a general relation between anomalous diffusion and anomalous heat conduction. In Sec III, we demonstrate the heat conduction and diffusion in several nonlinear lattices such as the FPU model and the FK model etc.. In Sec. IV, we discuss the heat conduction in quasi one dimensional lattice with transverse motion and the anomalous diffusion in a single walled nanotubes (SWNT). In Sec V, we give a detailed discussion on the heat conduction and diffusion in several billiard gas channels. In the last part, Sec VI, we give conclusions. 

\section{Anomalous diffusion and anomalous heat conduction}

The conclusion made by Prosen and Campbell\cite{Prosen00} states that in a momentum conserved one dimensional system, the thermal conductivity diverges with the system size $L$. However, the theory does not give any indication that how the thermal conductivity diverges with system size $L$.

Here we would like to find a microscopic origin of the
anomalous heat conduction observed in many 1D models. We shall not
restrict to any specific model. This should give us a more general
way to understand the heat conduction in 1D systems.

To establish a connection between the microscopic
process and the macroscopic heat conduction, let's consider a 1D
model of length $L$ whose two ends are put into contact with
thermal baths of temperature $T_L$ and $T_R$ for the left end and
the right end, respectively. Suppose the energy is transported by
energy carriers, phonons in lattices and particles in
billiard gas channels, from left heat bath to the right heat bath and
vice versa. The mean square of displacement of the carrier,
with velocity $v$, inside the system can be described by,
\begin{equation}
\langle \Delta x^2\rangle = 2Dv^{\alpha}t^{\alpha}.
\label{eq:Diffusion}
\end{equation}
The average time for the particle travels from the left end to the right end is
\begin{equation}
\langle t_{LR}\rangle\approx \langle t_{RL}\rangle \sim L^{2/\alpha}.
\label{tLR}
\end{equation}
We should stress that, the time given above is the averaged first passage time\cite{LW03,LW04}. It is taken over the distribution of the first passage time, or a truncated distribution of the first passage time if the average is not convergent like in the case of subdiffusion\cite{LW04} (see also Ref. \cite{Zaslavsky02}) for more regorous discussion.). Moreover, time (\ref{tLR}) is not the recurrence time because when the particle arrives at the opposite end its motion will be ``interrupted" due to of the interaction with the heat bath. It should be pointed out that in the cases we studied in this paper the exponent $\alpha$
in Eq. (\ref{eq:Diffusion}) and (\ref{tLR}) are same. 
 
The heat current, i.e. the energy change between the two heat baths in unit time, for a single particle moves from left to right and comes back is:
\begin{equation}
j_s=\frac{E_L-E_R}{\langle t_{LR}\rangle +\langle t_{RL}\rangle} \propto \frac{T_L-T_R}{\langle t_{LR}\rangle} =\frac{\Delta T}{\langle t_{LR}\rangle }.
\label{Heatcurrent}
\end{equation}
For $L$ particles in the system, the heat current is,
\begin{equation}
j=Lj_s \propto -\Delta T L^{1-2/\alpha},
\label{Jcurrent}
\end{equation}
then the heat conductivity is given by
\begin{equation}
\kappa =-j/\nabla T \propto L^{\beta} \label{kappa}
\end{equation}
with,
\begin{equation}
\beta=2-2/\alpha.
\label{alpha-beta}
\end{equation}

This relationship can be derived in a more rigorous way by using the fractional derivative\cite{LW03}. It connects heat conduction and diffusion
quantitatively. The main conclusion is that an anomalous
diffusion indicates an anomalous heat conduction with a divergent
(convergent) thermal conductivity. More precisely, our result
tells us that: a ballistic motion means thermal conductivity
proportional to the system size $L$, a normal diffusion means a
normal heat conduction obeying the Fourier law, a superdiffusion
means a divergent thermal conductivity, a subdiffusion means a
zero thermal conductivity in the thermodynamic limit. 

As the derivations of above equations are based on certain conditions, the relation Eq. (\ref{alpha-beta}) is not expected to apply to all anomalous transports. We believe that in other cases, where $\alpha$ in Eq. (\ref{tLR}) is not the same as that one in Eq. (\ref{eq:Diffusion}), $\alpha -\beta$ formula may be different from Eq. (\ref{alpha-beta}). In this sense, $\alpha$ in Eq.(\ref{tLR}) is the same $\alpha$ in Eq. (\ref{eq:Diffusion}) can be regarded as our pre-assumption, and our motivation is to investigate, under this assumption, what a connection between $\alpha$ and $\beta$ can be set up. 
The assumption that the $\alpha$ in both Eq. (\ref{eq:Diffusion}) and Eq. (\ref{tLR}) are the same one is based on the fractional differential equation's description of the diffusive process\cite{KlafterDing}, which is found to be useful in describing the anomalous transport in both chaotic and quasi-chaotic systems. However, to what extend, and how the fractional differential equations can be applied is  an interested topic and is still far from complete understanding. Our studies in Ref.\cite{LW03} and in this paper showed that at least in the examples we have studied (in certain parameter regimes), the prediction based on the fractional differential equation is in good agreement with the numerical simulations.

It is worth mentioning that, in past decades, a large amount of efforts have been devoted to the study of anomalous diffusion and its relationship with anomalous electric conductivity (see Ref. \cite{Bouchaud90} and the reference therein). So far, our work\cite{LW03} has been the first one to connect the anomalous diffusion with anomalous heat conduction. More importantly, recent years have witnessed renaissance of the research on anomalous diffusion and its connection with underlying dynamics. For more information, please refer to a recent extensive review\cite{Zaslavsky02} and the references therein.

In the following section, we investigate the diffusion in different nonlinear lattice models, in particular, the FPU model and the FK model, two representatives for nonlinear lattices without and with on-site potential, to see how the anomalous/normal diffusion is connected to the anomalous/normal conductivity.
We also check this relation in single walled nanotubes in Section IV and in the billiard gas channels in Section V.

\section{Anomalous heat conduction and diffusion in nonlinear lattice models}

To study diffusion in a lattice model, 
we first thermalize the system to an equilibrium state with temperature $T$ (corresponds to energy $E_0$), 
then the middle particle is given a much higher temperature $T_{\delta}$ (higher kinetic energy). The evolution of the energy profile along the chain is then recorded afterwards.  Quantitatively, the width of the pulse can be measured by its second moment

\begin{equation}
\sigma ^2(t)=\frac{\int (E(x,t)-E_0)(x-x_0)^2dx}{\int (E(x,t)-E_0)dx},
\label{eq:Sigma2t}
\end{equation}
where $E(x,t)$ is the energy distribution at time $t$. $x_0$ is the position of energy pulse at $t=0$. Generally, the energy profile spreads as 
\begin{equation}
\sigma ^2(t) =2Dt^\alpha, \quad (0<\alpha \leq 2).
\label{eq:Sigma2ta}
\end{equation}
To suppress statistical fluctuations, an ensemble average over
many realizations is performed when obtaining $\sigma^2(t)$.

Generally, one cannot define the same quantity for energy diffusion in lattice models as Eq.(\ref{eq:Diffusion}) like we do in billiard gas channels. In billiard gas channel, as we will show later on that $\langle \Delta x^2\rangle$ is calculated over the particle concentration while $\sigma^2$ in Eq. (\ref{eq:Sigma2t}) is calculated over the energy density distribution. In fact, in billiard gas channel, as the particle is the energy carrier, it can be shown that both $\langle \Delta x^2\rangle$ and $\sigma^2(t)$ are proportional to $t^{\alpha}$. However, for lattice model, the physical picture for the $\langle \Delta x^2\rangle$ is not clear, thus we use Eq. (\ref{eq:Sigma2t}) to describe the diffusion.

\subsection{The Fermi-Pasta-Ulam $\beta$ model}

As is well known that the FPU model plays an important role in the development of nonlinear science and the development of molecular dynamics (see the other  articles in this Focus issue). We should say that, the FPU model plays also an essential role in initiating the study of heat conduction in low dimensional systems in last years. In fact, at the start stage of the study in recent years, most works have been focused on the FPU model\cite{FPUKm,FPULepri,FHLZ98,HLZ00,Aoki01,FPUMode,FPUPer}.  The Hamiltonian of the Fermi-Pasta-Ulam $\beta$ model is
\begin{equation}
H=\sum_i \frac{p_i^2}{2m}+\frac{1}{2}(x_i-x_{i-1}-a)^2+\frac{\beta^*}{4}(x_i-x_{i-1}-a)^4
\label{FPUHaml}
\end{equation}
or 
\begin{equation}
H=\sum_i \frac{p_i^2}{2m}+\frac{1}{2}(q_i-q_{i-1})^2+\frac{\beta^*}{4}(q_i-q_{i-1})^4,
\label{FPUHam2}
\end{equation}
if the relative coordinates, $q_i (=x_i-ia)$, are used.

The heat conduction has been studied by different groups\cite{FPUKm,FPULepri,FHLZ98,HLZ00,Aoki01} numerically and found that $\kappa\sim N^{0.42\sim0.45}$. Both mode-coupling theory from Lepri\cite{FPUMode} and the Peierls equation study by Perezenev\cite{FPUPer} argued that the divergent exponent should be $2/5$ which is very close to the numerical values.
Here we give at the first time the diffusion behavior of the FPU model and try to testify the formula Eq. (\ref{alpha-beta}) in the FPU model.

\begin{figure}
\includegraphics[width=8.5cm]{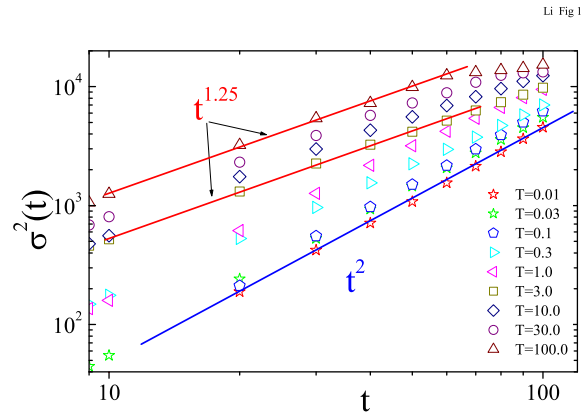}
\vspace{-.5cm}
\caption{\label{FPUsigma2t} 
$\sigma^2(t)$ versus time, $t$, for the FPU model at different temperatures. The temperature ranges from very low to very high. The chain length $N=501$. $10^4$ realizations are taken for the average. $T_{\delta}=10T$. It is clearly show the transition from ballistic motion (slope =2) at very low temperature to the anomalous diffusion at large temperature (slope =1.25). $\beta^*=1$. Note that the first few data for $T=0.1$ deviate from $t^2$ because the large fluctuations due to thermalization.}
\end{figure}

$\sigma^2(t)$ versus $t$ for different temperatures are shown in Fig.\ref{FPUsigma2t}. It is shown that the diffusion takes a smooth crossover from  ballistic one (with slope=2) at low temperature to an anomalous diffusion (slope =1.25) at high temperature. It is noted that at high temperature regimes ($T\ge 3.0$), the slope 1.25 does not depend on the temperature. It seems that this 1.25 is the high temperature limit. The slope 2 is the low temperature limit which is easy to understand as at very low temperature, the anharmonic term in the FPU model can be neglected.
We would like to point out that in doing the best fit for the slopes given in Fig.\ref{FPUsigma2t}, one needs to get rid of the boundary effects (same is true for Figs.3, 5 and 9). This can be easily down by observing the time when the front of the energy wave packet approaches the boundary. This time is used as the cutoff time for the best fit.

In fact, it is only a qualitative way to check when a wavepacket front approaches the boundary from the snapshot. The quantitative way is to use $\sigma^2(t)$. In the log-log scale $\sigma^2(t)$ plot, the $\sigma^2(t)$ increases with time $t$ until a critical time when the wave front hit the boundary and the reflection happens. After this critical time, the $\sigma^2(t)$ becomes saturate, some time even oscillating. 

If we insert $\alpha=1.25$ into formula (\ref{alpha-beta}), we obtain  $\beta=0.4$ which is exactly the same value obtained by mode-coupling method\cite{FPUMode} and the Peierls equation method\cite{FPUPer}, and very close to the numerical value, $\beta =0.42 \sim 0.45$,  from direct nonequilibrium thermal conductivity calculation.

\begin{figure}
\includegraphics[width=8.5cm]{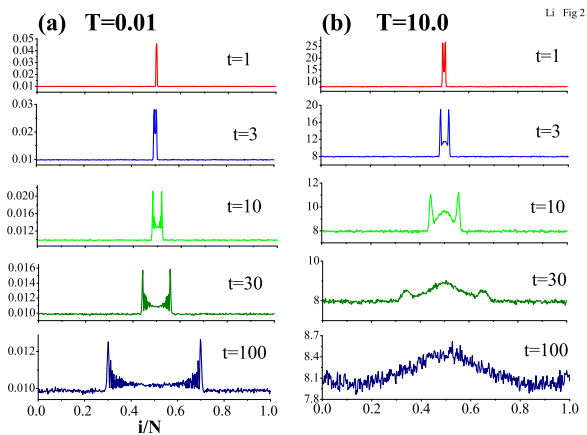}
\vspace{-.5cm}
\caption{\label{FPUsnapshot} Snapshots of the heat pulse spreading in the FPU model for different temperatures. (a) $T=0.01$; (b)$T=10$. In both cases, the chain is first thermalized to a state of temperature $T=0.01$ and $T=10$, then a heat pulse ($T_{\delta}=5 T$) is given to the middle particle. The chain length is $N=501$, and $\beta^*=1$.}
\end{figure}

Snapshots of two representative pulse propagation along the FPU lattice are shown in Fig. \ref{FPUsnapshot}(a) for low temperature and Fig. \ref{FPUsnapshot}(b) for high temperature. The difference of the diffusion in two different temperatures are clearly seen.

\subsection{The Frenkel-Kontorova model}

The FPU model discussed above is a representative lattice model without on-site potential. The Frenkel-Kontorova model, on the other hand, is a representative for a general one dimensional lattice with on-site potential. It describes a chain of particles connected by spring of constant $\gamma$, and put on the substrate. The average particle distance is $a$, and the periodic of on-site potential is $b$. The Hamiltonian of the
standard FK model is\cite{HLZ98}

\begin{equation}
 \tilde{H} =\sum \frac{P^2_i}{2m}+\frac{\gamma}{2}(X_i-X_{i-1}-a)^2-\frac{A}{(2\pi)^2}\cos \frac{2\pi X_i}{b}.
\label{eq:Ham0}
\end{equation}

Here $X_i$ is position of the i'th particle. For convenience of numerical calculations, the above Hamiltonian can be scaled to a dimensionless one \cite{HLZ98},

\begin{equation}
H =\sum \frac{p^2_i}{2m}+\frac{1}{2}(x_i-x_{i-1}-\mu)^2-\frac{V}{(2\pi)^2}\cos 2\pi x_i.
\label{eq:Ham1}
\end{equation}

After the scaling, the period of the on-site potential is unity, $\mu$ is the winding number. $\mu=1$, $m=1$, and the lattice constant is also taken to be unity in this paper. Of course, we can also express the Hamiltonian (\ref{eq:Ham0}) in terms of the displacement from the equilibrium, $q_i$. Substituting $x_i=i \mu+q_i$ into eq. (\ref{eq:Ham1}), we have,
$\tilde{H}=\sum \frac{p^2_i}{2m}+\frac{1}{2}(q_i-q_{i-1})^2-\frac{V}{(2\pi)^2}\cos (2\pi (i\mu +q_i)).$

\begin{figure}
\includegraphics[width=8.5cm]{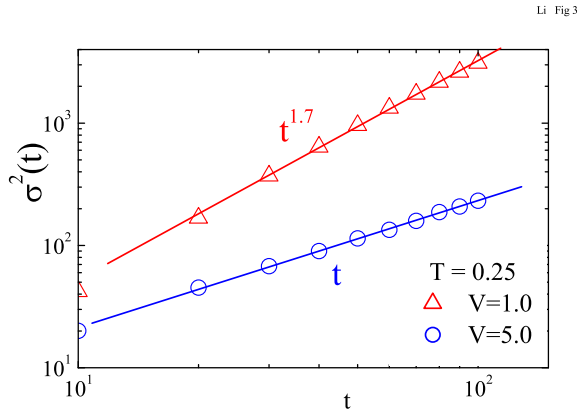}
\vspace{-.5cm}
\caption{\label{fig:FKsigma2t}$\sigma^2(t)$ versus time, $t$, for the FK model at different coupling constants. The system is thermalized at first at $T=0.25$. The chain length $N=201$. $5\times10^5$ realizations are taken for the average. At $V=1$, the energy transport superdiffusively, while at $V=5$ the energy transports diffusively which supports the results for normal heat conduction.}
\end{figure}

In Fig. \ref{fig:FKsigma2t} we show $\sigma ^2(t)$ versus $t$ in log-log scale such 
that the slope of the curve gives the value of 
$\alpha $. The temperature in both cases is $T=0.25$. It is clearly seen that the asymptotic slopes show different behaviors. In the case of large coupling constant, $V=5$, the energy transports diffusively, the heat conductivity is a system-size independent constant, as we observed before\cite{HLZ98}. 
However in the case of small coupling constant, $V=1, \sigma^2(t)\sim t^{1.7}$, the energy transports superdifussively.   According to Eq.(\ref{alpha-beta}), this corresponds to a divergent thermal conductivity which agrees qualitatively with the results from Savin and Gendelman\cite{FKSG03} 

The snapshots also show complete different story for $V=1$ and $V=5$ cases.
In the former case, the heat pulse splits into two pulses and propagates along two opposite directions. However, in the case $V=5$ the pulse never splits, it only becomes fatter and fatter. 

As is well known that the ground state of the Frenkel-Kontoroval model can be cast into a standard map. Then there exists a one-to-one relationship between these two models (see review article\cite{FKreview}). For example, an incommensurate state corresponds to an invariant curve with an irrational winding number in the phase space and a commensurate state corresponds to an invariant curve with rational winding number. In the standard map, the anomalous diffusion in both $x$ and $p$ directions has been extensively studied by Zaslavsky et al\cite{Mapdif}. It is still not clear and deserve further investigation how the anomalous diffusion in the standard map can be connected to the anomalous diffusion in the FK model observed in Fig.(\ref{fig:FKsigma2t}a).

\begin{figure}
\includegraphics[width=8.5cm]{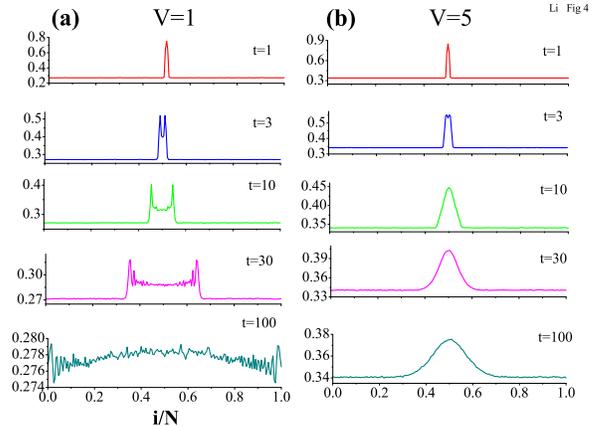}
\vspace{-.5cm}
\caption{\label{FKsnapshot}Snapshots of the heat pulse spreading in the FK model at different coupling constant. (a) $V=1$; (b)$V=5$. The other conditions are the same as that ones in Fig.\ref{fig:FKsigma2t}. }
\end{figure}

\subsection{The $\phi^4$ model}

Like the FK model, the $\phi^4$ model has an on-site potential, its (dimensionless) Hamiltonian is\cite{HLZ00,Aoki00}
\begin{equation}
H=\sum_i \frac{p_i^2}{2m}+\frac{1}{2}(q_i-q_{i-1})^2+\frac{\beta^*}{4}q_i^4.
\label{hamphi4}
\end{equation}

\begin{figure}
\includegraphics[width=8.5cm]{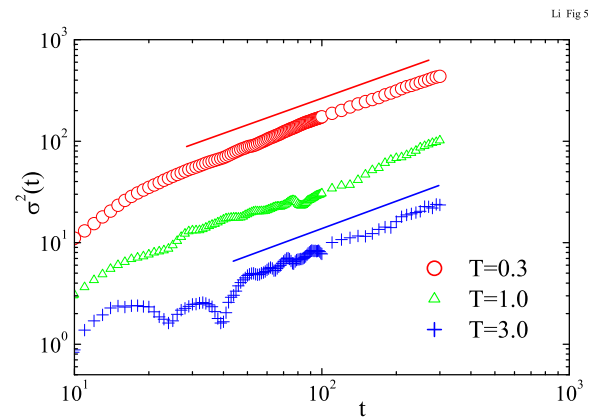}
\vspace{-.5cm}
\caption{\label{fig:FFsigmat}The $\sigma^2(t)$ versus time, $t$, for the $\phi^4$ model at different temperature with $\beta^*=10$. In all three cases, $\sigma^2(t) \sim t$ asymptotically.}
\end{figure}

Due to the presence of the on-site potential that breaks down the momentum conservation, the heat conduction in this model has been shown to obey the Fourier law\cite{HLZ00,Aoki00}. 
We show here $\sigma^2(t)$ versus $t$ for different temperatures in Fig \ref{fig:FFsigmat}. The slopes in all three cases are approximately one, thus the vibration energy transports diffusively, that supports the conclusion of finite thermal conductivity.

\begin{figure}
\includegraphics[width=8.5cm]{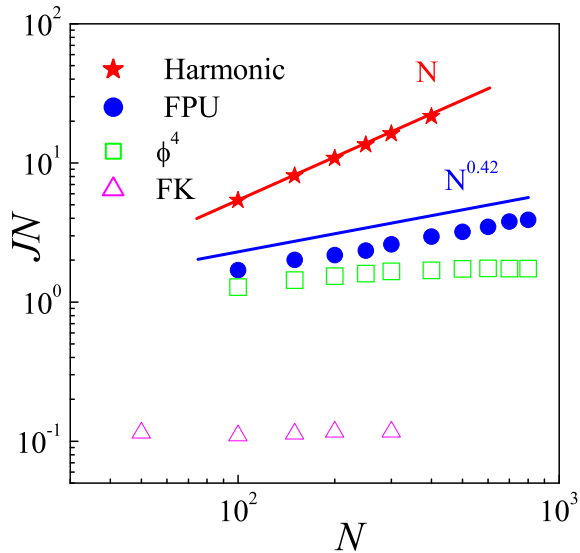}
\vspace{-.5cm}
\caption{\label{kappa-summary} $JN$ (proportional to thermal conductivity $\kappa$) versus system size $N$ for a harmonic chain ($\star$), the FPU model ($\bullet$), the $\phi^4$ model ($\square$) and the FK model ($\triangle$). The $JN$ diverges as $N$ in the harmonic model and diverges as $N^{0.42}$ in the FPU model, whereas it is constant in the $\phi^4$ model and the FK model. The data for the FK model are taken from Ref.\cite{HLZ98}, while the others from Ref\cite{HLZ00}.}
\end{figure}

For comparison, we summarize the $JN$ (proportional to thermal conductivity) for different models in Fig \ref{kappa-summary}.

\subsection{Coupled rotator model}

Another interesting lattice model is the so-called coupled rotator model. 
The Hamiltonian  is:
\begin{equation}
H=\sum_i \frac{p_i^2}{2m}+(1-\cos(x_i-x_{i-1}-a)),
\label{hamrotator}
\end{equation}
where $a$ is the lattice constant which is taken to be $2\pi$. Obviously, due to the absence of the on-site potential, the momentum is conserved in this model, nevertheless, 
numerical results\cite{Rotator1,Rotator2} show that the heat conduction obeys the Fourier law. Recently, Flach \cite{Rotator3} argues that the finite thermal conductivity might be a numerical phenomenon, the thermal conductivity of this model is divergent. He conjectures that in order to observe the divergent thermal conductivity, one needs to go to very large system size, so large that it may exceed today's computer facility. In the following, we shall clarify this puzzle. 

Generally, heat flux in a 1D lattice is defined as:
\begin{equation}
J = \frac{1}{N} \left( \sum_{i=1}^{N} f_i \dot{x}_i 
 + \sum_{i=1}^{N} \frac{1}{2a}m \dot{x}_i^2\dot{x}_i\right)=J_1+J_2,
\end{equation}  
where $f_i$ is the interaction force between particle $i$ and $i+1$. In this model, 
$f_i=\sin(x_i-x_{i+1}-a)$. $J_1$ corresponds to energy flux transfered from one particle to the other, while $J_2$ the energy flux due to the particle's movement.
Usually, in other lattice models such as the FPU model, the FK model and etc, $\langle J_2\rangle $ vanishes because the particle vibrates only around its equilibrium position. However for coupled rotator model, things are different. At lower temperature, so low that the kinetic energy is not large enough to make the rotator rotating, $\langle J_2\rangle $ also vanishes. However, if the temperature is large enough so that the rotator can overcome the energy barrier to start rotating, then, $\langle J_2\rangle $ is not zero. This part of energy current does not go through the system from high temperature end to the low temperature end. If we use the non-equilibrium method to calculate the thermal conductivity, for example, by puttiny thermal bathes on the two ends of the lattice, then $\langle J_2\rangle$ does not contribute to the thermal conductivity. 

\begin{figure}
\includegraphics[width=8.5cm]{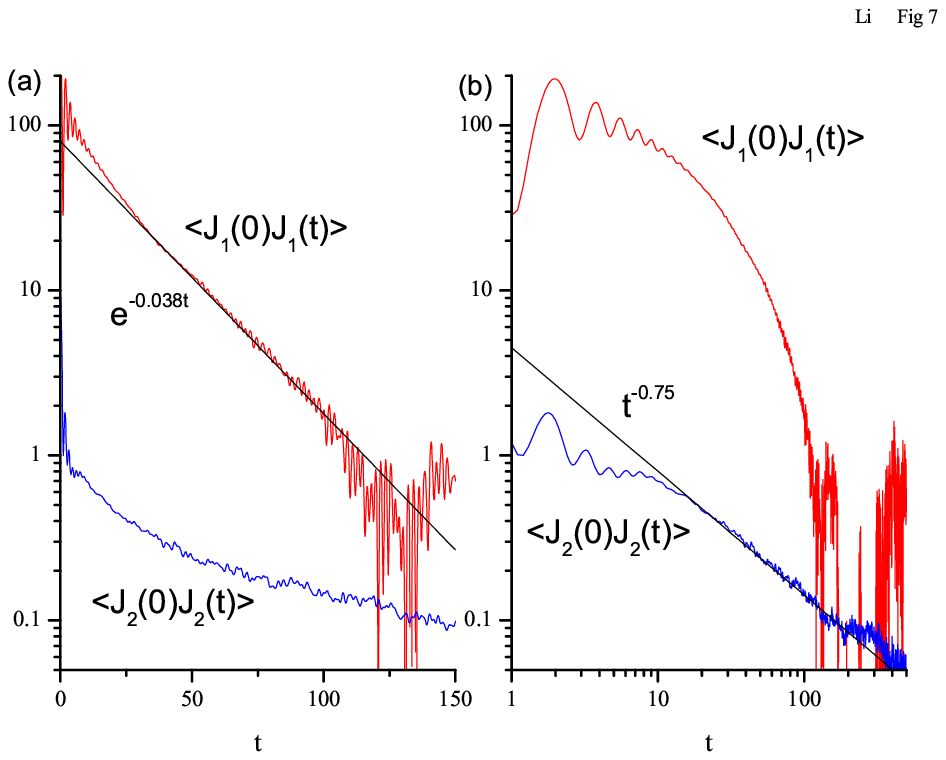}
\vspace{-.5cm}
\caption{\label{fig:rotakubo} Current-current correlation for 2000 rotators versus time. (a) in log-linear scale, (b) in log-log scale. 
 }
\end{figure}

When one uses the Green-Kubo formula to calculate the thermal conductivity, the situation becomes different.
 In Fig. \ref{fig:rotakubo}, we show J-J correlation contributed by $J_1$ and $J_2$. 
$\langle J(0)J(t)\rangle =\langle(J_1(0)+J_2(0))(J_1(t)+J_2(t))\rangle=\langle J_1(0)J_1(t)\rangle +\langle J_2(0)J_2(t)\rangle+ \langle J_1(0)J_2(t)\rangle+\langle J_2(0)J_1(t)\rangle$.
The ensemble average of the third and fourth terms vanish because, $J_1$ is symmetric with space reflection, while $J_2$ anti-symmetric.  Thus $\langle J(0)J(t)\rangle =\langle J_1(0)J_1(t)\rangle+\langle J_2(0)J_2(t)\rangle$. Clear exponential decay and power-law decay can be seen for $\langle J_1(0)J_1(t)\rangle$ \cite{Rotator1} and $\langle J_2(0)J_2(t)\rangle$, respectively. This implies that if only $J_1$ is taken into account, then one has a normal heat conduction as is observed by previous numerical simulations\cite{Rotator1,Rotator2}. Otherwise if both are considered, then the J-J correlation decays slower than $t^{-1}$, as a result, the thermal conductivity diverges as Flach conjectured.

\section{Lattice model with transverse motion and single walled carbon nanotubes}

\subsection{A polymer like quasi one dimensional lattice}

To study the cross over of the divergent exponent from $2/5$ to $1/3$, Wang and Li\cite{WL04} introduced a polymer like chain. The chain consists of $N$ point particles
with mass $m$ on a 1D lattice. The particles have both longitudinal and transverse motions.
The lattice fixes the connectivity
topology such that only the neighboring particles interact.  The
Hamiltonian is given by
\begin{eqnarray}
H({\bf p}, {\bf r}) & = & \sum_{i} \frac{{\bf p}_i^2}{2m} +   
\frac{1}{2} K_r \sum_{i} \Bigl( | {\bf r}_{i+1}-{\bf r}_i | -a\Bigr)^2 
\nonumber
\\
&& +\, K_\phi \sum_{i} \cos \phi_i,
\end{eqnarray} 
where the position vector ${\bf r}=(x,y)$ and momentum vector ${\bf p}
=(p_x, p_y)$ are two-dimensional; $a$ is lattice constant.  
The minimum energy state is at $(ia,0)$ for $i=0$ to $N-1$.
If the
system is restricted to $y_i=0$, it is essentially a 1D gas with
harmonic interaction.  The coupling $K_r$ is the spring constant;
$K_\phi$ signifies bending or flexibility of the chain, while $\phi_i$
is the bond angle formed with two neighbor sites, $\cos
\phi_i = - {\bf n}_{i-1} \cdot {\bf n}_{i}$, and unit vector ${\bf n}_i = \Delta
{\bf r}_i /| \Delta {\bf r}_i|$, $\Delta {\bf r}_i = {\bf r}_{i+1} -
{\bf r}_{i}$. 

\begin{figure}
\includegraphics[width=8.5cm]{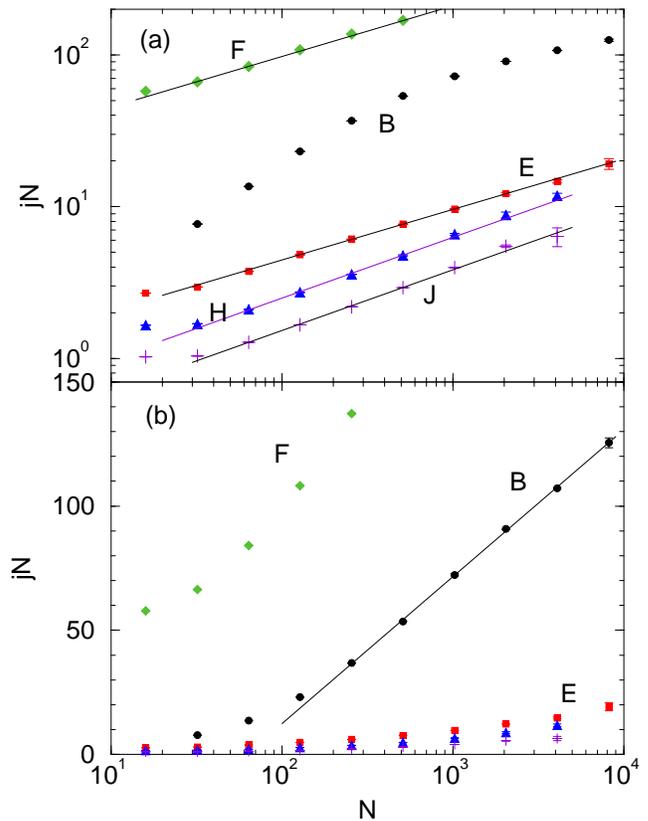}
\caption{\label{Polymerkappa} $jN$ vs $N$ on double-logarithmic (a)
and linear-log plot (b). The parameters  ($K_\phi, T_L, T_H$) of the model 
are,
set B: (1, 0.2, 0.4),
set E: (0.3, 0.3, 0.5),
set F: (1, 5, 7),
set H: (0, 0.3, 0.5),
set J: (0.05, 0.1, 0.2).  All of them have
$K_r = 1$, mass $m=1$, lattice spacing $a=2$.  The straight lines on F and E have
slope 1/3, while the slope on H and J is 2/5. This picture is taken from Ref\cite{WL04}.}
\end{figure}

The conductivity results, $jN$, is shown in Fig. \ref{Polymerkappa}.
It is shown that at the high temperature the thermal conductivity diverges with system size $L$ as, $L^{1/3}$, whereas at the low temperature regime, the thermal conductivity diverges with power $2/5$. 
A detailed mode-coupling theory analysis\cite{WL04} tells us that at high temperature the transverse modes are strongly coupled with the longitudinal ones while at the low temperature regime the longitudinal modes are weakly coupled or completely uncoupled with the transverse mode. This explains qualitatively and quantitatively the crossover from $2/5$ to $1/3$. More interestingly, in an intermediate regime we also observe the $\ln N$ divergent thermal conductivity for which a theory is still lacking.

\subsection{Single walled carbon nanotubes}

Carbon nanotube is one of exciting nano-scale materials discovered in recent years. It reveals many excellent mechanical, thermal and electronic properties\cite{textbook}. However, compared with  mechanical and electronic properties, much less is known about thermal conductivity in nanotubes. 
Although some researches have been done\cite{Nanoheat}, many important and fundamental questions remain unsolved. For example, does heat conduction in a SWNT obey the Fourier law? If it does not, then, how does the thermal conductivity diverge with the tube length $L$?  Is heat conduction in carbon nanotubes like that one in 1D lattice model or 2D lattice model or 3D bulk material?

The Hamiltonian of the carbon SWNT is: 
\begin{equation}
H=\sum_i \left(\frac{p_i^2}{2m_i}+V_i\right),\quad V_i=\frac{1}{2}\sum_{j,j\neq i}V_{ij}
\label{SWNTHam}
\end{equation}
where 
$V_{ij}=f_c(r_{ij})[V_R(r_{ij})+b_{ij}V_A(r_{ij})]$ is the Tersoff empirical bond order potential. $r_{ij}$ is the distance between the $i$'th and the $j$'th atom. $
V_R(r_{ij})=A\exp (-\lambda r_{ij})$, and $V_A(r_{ij})=-B\exp (-\mu
r_{ij})$ are the repulsive and attractive parts of the potential,
and $ f_c(r)=1$, for $r<R; (1+\cos \frac{\pi (r-R)}{S-R})/2,$ for $ R\le r\le S $ and $0,  r>S$. $b_{ij}$ are the so-called bond parameters
depending on the bounding environment around atoms $i$ and $j$, 
they implicitly contain many-body information. $A, B, \lambda, \mu, R$ and $S$ are parameters. For detailed information please refer to
Ref.\cite{Tersoff}. For the comparison we also study one dimensional carbon lattice with the same interaction, namely, $
V_{i,i-1}=f_c(x_{i,i-1})[V_R(x_{i,i-1})+V_A(x_{i,i-1})].
$

Molecular dynamics simulation\cite{ZL04} show that at low temperature such as $2K$, for both
the SWNT and the 1D carbon lattice, there is no temperature gradient. It resembles
the 1D lattice model with harmonic interaction
potential\cite{Lebowitz}. This can be understood from the Taylor
expansion of the Tersoff potential by keeping up to the second order term.
Because at low temperature, the vibrations of atoms are very small, the
potential can be approximated by a harmonic one, and the vibration
displacement in transverse direction is much smaller than the one along the
tube axis and can be neglected.  This result means that the phonons transport
ballistically in the SWNTs at low temperature. 

However at room temperature, the situation changes dramatically. At $300$K there is still no temperature gradient in 1D
carbon lattice, but in the SWNT, temperature gradient is set up. In 1D carbon lattice
with Tersoff potential, the increase of temperature does not change the harmonic
character because the strength of Tersoff potential; while in the SWNT at room temperature, there
are anharmonic terms due to the transverse vibrations. 

\begin{figure}
\includegraphics[width=8.5cm]{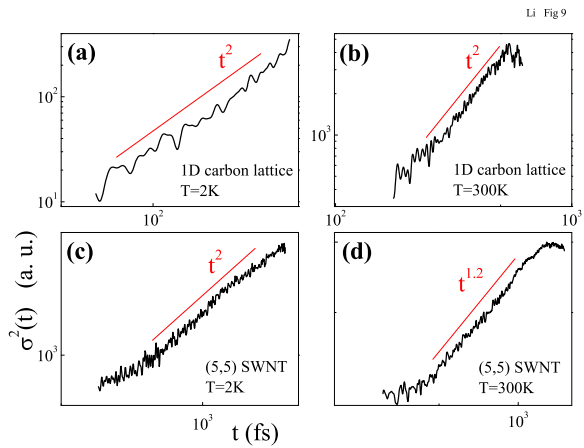}
\vspace{-.5cm}
\caption{$\sigma^2(t)$ versus time. (a) 1D lattice at $2K$; (b) 1D lattice at 300K; (c)(5,5) SWNT at 2K; (d) (5,5) SWNT at 300K. 
$10^3$ realizations are performed to do the average.}
\label{fig:StandDev}
\end{figure}

In Fig. \ref{fig:StandDev} we show $\sigma ^2(t)$ versus $t$ in double logarithmic scale, so
that the slope of the curve before the turning point (due to boundary reflection) gives the value of 
$\alpha $. Fig.\ref{fig:StandDev}(a) and (b) show the results for the 1D lattice at temperature 2K and 300K, respectively. It is clearly seen that the slopes are 2 in both
cases. This means that energy transports ballistically. However, for the SWNT, the
situation is different. At T=2K, energy transports
ballistically (Fig. \ref{fig:StandDev}(c)) like in the 1d carbon lattice, whereas at T=300K, energy transports superdiffusively with $\alpha \approx 1.2$ (Fig.\ref{fig:StandDev}(d)). If we use $\alpha\approx 1.2$, then, according to our formula Eq. (\ref{alpha-beta}), we have $\beta \approx 1/3$ which agrees very well with the result, $\beta \approx 0.32$ obtained by Maruyama\cite{Maruyama} with non-equilibrium molecular dynamics method.

\begin{figure}
\includegraphics[width=8.5cm]{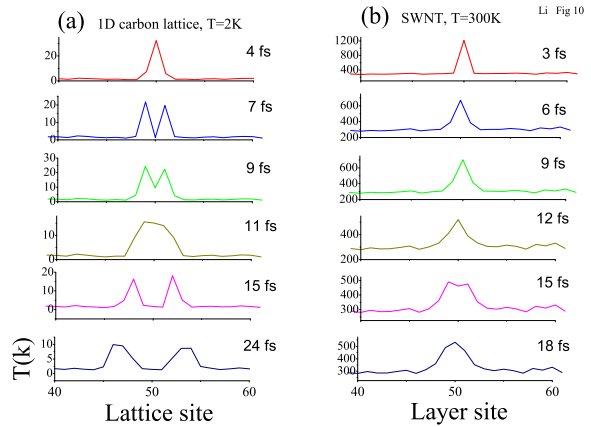}
\vspace{-.5cm}\\
\caption{Snapshots of heat pulse spreading in a $1D$ carbon lattice and a (5,5) SWNT at
different temperatures. (a) $1D$ lattice at $2K$. (b) The SWNT at $300K$.
For 1D lattice, at T=300K, the pulse spreads in the same manner and same speed as for T=2K, whereas for SWNT, the pulse spreads faster at T=300K than at T=2K. For quantitative comparison, please refer to Fig (\ref{fig:StandDev}). }
\label{fig:Pulse}
\end{figure}

 Snapshots of two representative pulse propagation along the tube/lattice are shown in Fig \ref{fig:Pulse}. In 1D carbon lattice, as demonstrated by Fig\ref{fig:Pulse}(a), both at T=2K and T=300K, a peak splits into two peaks very quickly, and each peak spreads ballistically to the ends of the lattice.  However, in the SWNT, the pulse spreads out, but with different rate at 2K and 300K. Quantitatively, the width of the pulse can be measured by its second moment as is given by Fig \ref{fig:StandDev}.

It is worth pointing out that, in many
experiments or practical applications, the SWNT/lattice is put
on a substrate, then one should add one additional potential term in 
Hamiltonian (\ref{SWNTHam}). With introduction of the on-site
potential, heat conduction and energy diffusion are expected to be
normal like that one in the 1D FK model\cite{HLZ98} and $\phi^4$ model\cite{HLZ00,Aoki00}. 

\section{Billiard gas channels} 

In this section, we give a brief review on recent years' studies on heat conduction in billiard gas channels and show our study on the diffusion. In fact, in addition to heat conduction problem\cite{Alonso99,LWH02,LCW03,Alonso02,LCWP04}, billiard models have been used to study other macroscopic transport\cite{Lorenz05,Chapman70,Hauge74,Hvan82,Gaspard99,Dettmann00}.
For example, the Lorentz gas model  has been used for the study of connection between the electrical conductivity and diffusion\cite{Chernov}. Moreover, in recent years, extensive studies on the chaotic dynamics and the origin of statistical law has been done by using different types of billiard gas model\cite{Zaslavsky02,Zaslavsky99}.

\begin{figure}
\includegraphics[width=8.5cm]{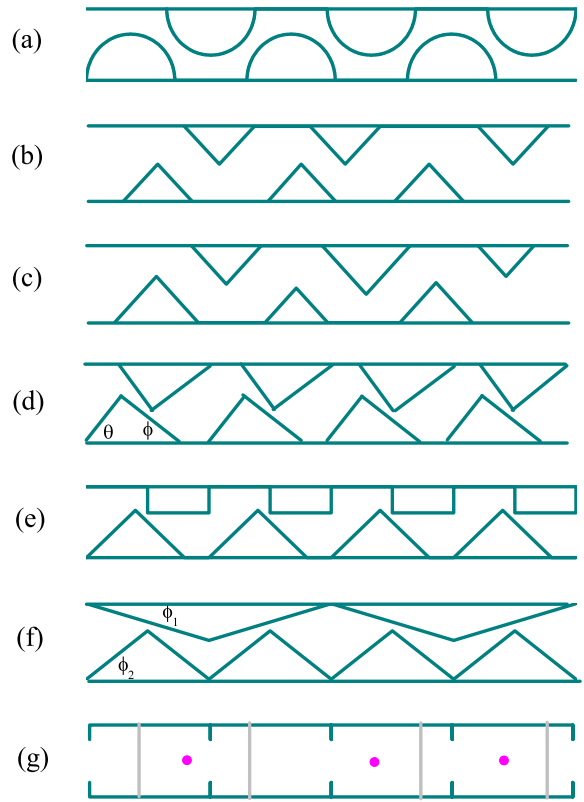}
\vspace{-.2cm}
\caption{\label{fig:billiard}The geometrical configuration of billiard gas channels.
(a) The Lorentz gas channel\cite{Alonso99}. (b) The disordered (height) Ehrenfest channel\cite{LWH02}. (c) The disordered (position) Ehrenfest channel\cite{LWH02}. (d) The triangle gas channel\cite{LCW03}. (e) The triangle-square channel. (f) The polygonal gas channel\cite{Alonso02}. (g) The alternate mass hard-core potential model\cite{LCWP04}. }
\end{figure}

In the study of heat conduction, the two ends of the billiard gas channel (see Fig\ref{fig:billiard}) are put into contact with heat baths. The heat baths are modeled by stochastic kernels
of Gaussian type,  namely, the probability distribution of
velocities for particles coming out from the baths is

\begin{eqnarray}
P(v_x) =
\frac{|v_x|}{T}\exp\left(-\frac{v^2_x}{2T}\right),\nonumber\\
P(v_y) = \frac{1}{\sqrt{2\pi T}}\exp\left(-\frac{v^2_y}{2T}\right)
\label{Gaussian}
\end{eqnarray}
for $v_x$ and $v_y$, respectively.

The temperature field at a stationary state is calculated by time averages by dividing the configuration space into a set of boxes ${C_i}$ \cite{Alonso99}. The time spent within a box in the $j$th visit is denoted by $t_j$ and the total  number of crossings of a box $C_j$ during the simulation is $M$. The  temperature field is defined by\cite{Alonso99}
\begin{equation}
T_{C_i} = \langle E \rangle_{C_i} = \frac{\sum_j^M t_jE_j(C_i)}{\sum_j^M t_j}.
\end{equation}
Then it is projected on $x$ direction (the transport direction). Here $E_j(C_i)$ is the kinetic energy of the particle at its $j$'th crossing of the $C_i$ box.
The heat flux is calculated by the change of energy carried through to the left and right ends by the particles,
\begin{equation}
J = \frac{1}{t_M}\sum_{j=1}^{M} \Delta E_ j,
\end{equation}
where $\Delta E_j= (E_{in} - E_{out})_j$ is the energy change at the $j$th collision with a heat bath, $t_M$ is the total time spent for $M$ such collisions.

In numerical simulation, we compute the flux for a single particle $J_1$. The scaled heat flux is $J_N(N) = NJ_1(N)$\cite{Alonso99}, where $N$ is the number of the cells. Each cell has length $a$, thus the channel has length $L=Na$.

The diffusive property of the particles in the channel is quantified by the mean square displacement $\langle \Delta x^2 (t)\rangle =\langle (x(t) - x(0))^2 \rangle$. Here $\langle\cdots \rangle$ means the ensemble average over many particles.

In the following we discuss the heat conduction and diffusion properties in the Lorentz gas channel\cite{Alonso99}, the Ehrenfest gas channel\cite{LWH02}, the triangle channel\cite{LCW03}, the polygonal billiard channel\cite{Alonso02}, and the alternate mass hard-core potential model\cite{LCWP04}, respectively.

\subsection{The Lorentz gas channel}

In order to answer the question of whether the exponential instability plays a crucial role in the Fourier law, Alonso et al\cite{Alonso99} studied the heat conduction in the Lorentz gas channel (see Fig \ref{fig:billiard}(a) for its configuration). 
The model consists of two
parallel lines of length L at distance H, that is taken to be
unity, and a series of semicircles of radius R placed in a
triangular lattice along the channel. By construction no
particle can move along the horizontal direction without
colliding with the disks. The dynamics in the Loretnz gas is
mixing and all trajectories with nonzero projection on the
$x$ direction are of hyperbolic type; further it has positive
Kolmogorov-Sinai entropy and a well defined diffusion\cite{Bunimovich}. It has been demonstrated\cite{Alonso99} that the heat conduction in this model obeys the Fourier law.

\subsection{The Ehrenfest gas channel}

Instead of the semicircles of radius R, we place right triangles along the chain, namely, in each cell, we have two triangles, one on the bottom wall, the other on the top. The triangles are placed at the position of $x=1,3,\cdots$ (arbitrary unit). The height of channel is $H=1.1$, while the height of the triangle is $0.6$. The unit cell length is 4.
The channel of length $N$ is $N$ repetitions of the cell. Due to the existence of free path, this model demonstrate a superdiffusive motion, i.e. $\sigma^2\equiv\langle \Delta x^2\rangle \propto 2Dt^{\alpha}$, with $\alpha \approx 1.672\pm 0.003$ \cite{LWH02}.
The heat conductivity diverges with the system size as $\kappa \propto N^{0.814}$. This $\beta=0.814$ is very close to that value 0.804 obtained from the formula eq.(\ref{alpha-beta}) by inserting $\alpha=1.672$.

\begin{figure}
\includegraphics[width=8.5cm]{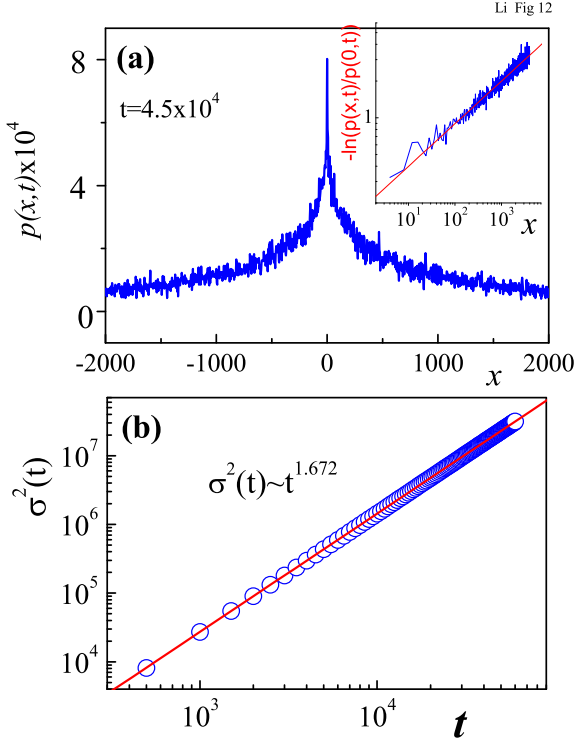}
\vspace{-.2cm}
\caption{Energy diffusion in the Ehrenfest gas channel. (a) $P(x,t)$ versus $x$ for a fixed $t=45,000$. Inset of (a), we show the scaled PDF in logarithmic scale, $-\log (P(x,t)/P(x,0))$. (b) $\sigma^2(t)$ versus time $t$.}
\label{fig:Ehren-df}
\end{figure}

In our numerical simulations, initially the particles are put at $x=0$ (the middle of the channel), with a uniform concentration distribution along the $y$ direction of the channel. Their speeds are set to be unit with a uniformly distributed direction in $x-y$ plane. Numerically, it is found that probability distribution function (PDF) (the probability of finding particle at position $x$ at time $t$) can be described by 
\begin{equation}
P(x,t)=P(0,t) \exp\left(-c_0|x/\sigma(t)|^{\gamma}\right),
\label{eq:SuperDfPDF}
\end{equation}
with $c_0=1.58$, and $\gamma=0.351$.
In Fig. \ref{fig:Ehren-df}, $P(x,t)$ is drawn for $t=45,000$, in the insert we draw $-\log (P(x,t)/P(0,t))$ versus $x$ which shows a good power law. 
As we shall demonstrate later on, this formula (\ref{eq:SuperDfPDF}) seems to be quite general for all superdiffusive case in billiard channels studied in this paper. The only difference lies in $c_0$, $\gamma$ and $\sigma(t)$.

{\bf Channel with triangles at random positions} 

In order to obtain a diffusion motion in the Ehrenfest gas channel, we modified above periodic structure by introducing disorder. The first way to do so is to make the positions of the triangles random (Fig. \ref{fig:billiard}(b)), namely, $x_i=d*R_i$, where $x_i$ is the position away from the periodic structure.  

{\bf Channel with right triangles of random heights}

The second way is to let the heights of the triangles random, namely,
\begin{equation}
h_i = h_0 + d*R_i,\quad i=1,2,\cdots,2N
\end{equation}
where $\{R_i\}$ are random numbers uniformly distributed in the interval $[-1,1]$, $d$ is the magnitude of disorder. $h_i<H$, where $H$ is the width of the channel. The geometry is shown in Figure \ref{fig:billiard}(c) shows the geometry of the channel.  We take $H=1.1$, $h_0=0.6$, and $d \in [0,0.4]$. 

Our results\cite{LWH02} shown that in both disordered Ehrenfest channels, the heat conduction obeys the Fourier law due to the normal diffusion induced by the disorder.

\subsection{The triangle gas channel}
In order to study further whether the exponential instability is a necessary condition as proposed in Ref.\cite{Casati84}, we consider  a two dimensional billiard model which
consists of two parallel lines of length $L$  at distance $H$  and
a series of triangular scatterers, Fig.\ref{fig:billiard}(d). In this geometry, no
particle can move between the two reservoirs without suffering
elastic collisions with the triangles. Therefore this model is
analogous to the one studied in Lorentz gas channel with triangles
instead of discs and the essential difference is that in the
triangular model discussed here the dynamical instability is
linear and therefore the Lyapounov exponent is zero. 

The numerical results from this model depends sensitively on whether the ratio $\theta/\pi$ and $\phi/\pi$ are rational or irrational numbers. In the case of irrational ratios, the system shows normal diffusion and the heat conduction obeys the Fourier law, while in the case of rational rations, the system shows a superdiffusive behavior, $\langle \Delta x^2\rangle=2Dt^{1.178}$ as shown in Fig. 6 in Ref.\cite{LCW03}. The heat conductivity diverges with the system size as $\kappa\approx N^{0.25\pm 0.01}$ by using larger system size. Again, by substituting $\alpha=1.178$ into formula (\ref{alpha-beta}), one gets, $\beta=0.30$ which is close to $0.25$. The deviation is due to the finite size effect\cite{LW03}.

\subsection{The triangle-square gas channel}

The triangle-square gas channel (see Fig\ref{fig:billiard}(e) for its geometry) consists of isosceles rectangular triangles on the bottom and the rectangles on the top.
The length of the fundamental cell is one, the height of the channel is $H=0.6$. The 
height of the isosceles rectangular triangle is 0.4, and the
width and the height of the rectangle is 0.25 and 0.5, respectively. 

\begin{figure}
\includegraphics[width=8.5cm]{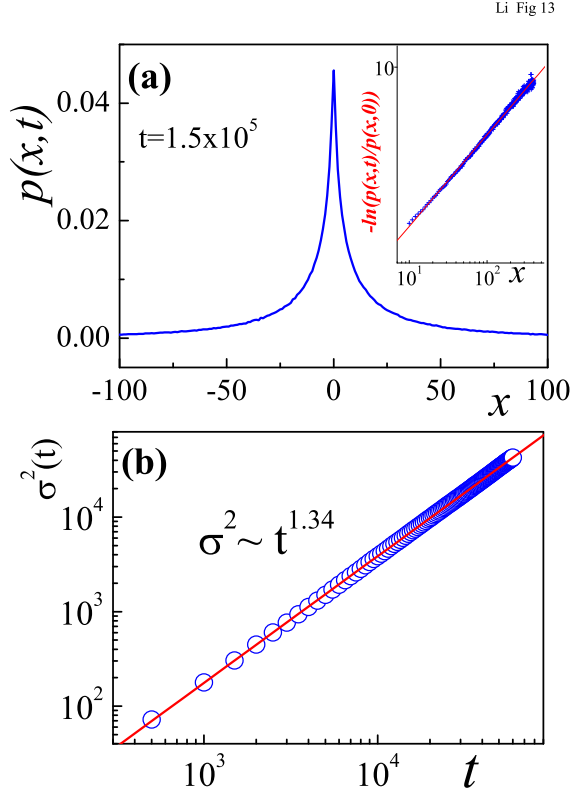}
\vspace{-.5cm}
\caption{Energy diffusion of the triangle-square gas channel.}
\label{fig:tri-rect-df}
\end{figure}

The particle's motion inside the channel is found to be superdiffusive.
Surprisingly, we also found that the PDF can be described by Eq.(\ref{eq:SuperDfPDF})
with $c_0=1.95, \gamma=0.376$ and,
$\sigma^2(t) \sim t^{1.34}$, see Fig.\ref{fig:tri-rect-df}. The $\beta$ value calculated from the diffusion is $\beta=2-2/1.34=0.51$. This is not far away from $\beta=0.46$, the direct non-equilibrium calculation by putting two heat baths on the two ends of the channel.

\subsection{The polygonal gas channel}

Another interesting billiard gas channel was proposed by Alonso et al\cite{Alonso02}. The geometry of this model is shown in Fig\ref{fig:billiard}(f). The height of the channel is $H=1$. In a fundamental cell, there are two isosceles triangle with angle $\phi_2$ on the bottom and a isosceles triangle with $\phi_1$ on the top. The cell length along the horizontal direction is 2d, $d=h/(\tan \phi_1+\tan\phi_2/2)$. 

By fixing $\phi_1=(\sqrt{5}-1)\pi/8$, and adjusting the angle $\phi_2=\pi/q$, with $q=3,4,5,6,7,8,9$, this model displays a very interesting transition from a superdiffusive motion ($q=3,5,6,7$), to a subdiffusive motion ($q=4$), and normal diffusion at $q=8$ and $9$.
The subdiffusive motion is quite interesting for heat conductivity, as $\alpha<1$ in this case, then according to our formula (\ref{alpha-beta}), we immediately have $\beta<0$, which means that the thermal conductivity of such subdiffusive system is an insulator in thermodynamic limit. 

In the subdiffusive case, Alonso et al\cite{Alonso02} shown that 
 $\alpha=0.86$, and the thermal conductivity
goes as $\kappa \sim L^{-0.63}$.  According to
our formula (\ref{alpha-beta}), if $\alpha=0.86$, $\beta =-0.33$
which is larger than the one obtained by Alonso et
al\cite{Alonso02}. This is not a surprise, because the channel
length in their study of thermal conductivity is too small ($L\le
40$). If the channel is longer, the value of $\beta$ will become
much more closer to our theoretical estimation ($\beta=-0.33$). To
demonstrate this, we extend the thermal conductivity simulation
from $L\in[1,40]$ used by Alonso et al \cite{Alonso02} to
$L\in[40,80]$, and find that $\beta=-0.48$ 
which is more closer to $\beta=-0.33$ than the one obtained by
Alonso et al. If $L \to \infty$, one can expect $\beta$ goes to
$-0.33$ eventually.

\subsection{The alternate mass hard-core potential chain}

The billiarde gas models discussed so far do not have local thermal equilibrium, thus people may argue about the validity of temperature definition\cite{Dhar299}. In order to overcome this problem, we have recently introduced a gas channel that has local thermal equilibrium, and also exhibits normal diffusion, but with zero Lyapunov exponent\cite{LCWP04}. 

The model is identical to the alternate mass hard-point gas \cite{2mass}, namely it consists of a 
one-dimensional chain of elastically colliding particles with alternate masses $m$ and $M$.
Here, however, in order
to prevent total momentum conservation we confine the motion of particles of mass $M$ (bars) inside unit cells of size $l=1$. 
Schematically the model is shown in Fig. \ref{fig:billiard}(g) in which particles with mass $m$ move horizontally and collide with bars of 
mass $M$ which, besides suffering collisions with the particles, are elastically reflected back at the edges of their cells. 
In between collisions, particles and bars move freely.
The total length of the system is $L= Nl$ where $N$ is the number of fundamental cells. 

Our numerical results clearly indicate that our model, contrary to the
translationally invariant model\cite{2mass}, obeys the Fourier law. 
The only difference between the two models is total momentum conservation. Our results support the proof by Prosen and Campbell\cite{Prosen00}.

\begin{figure}
\includegraphics[width=8.5cm]{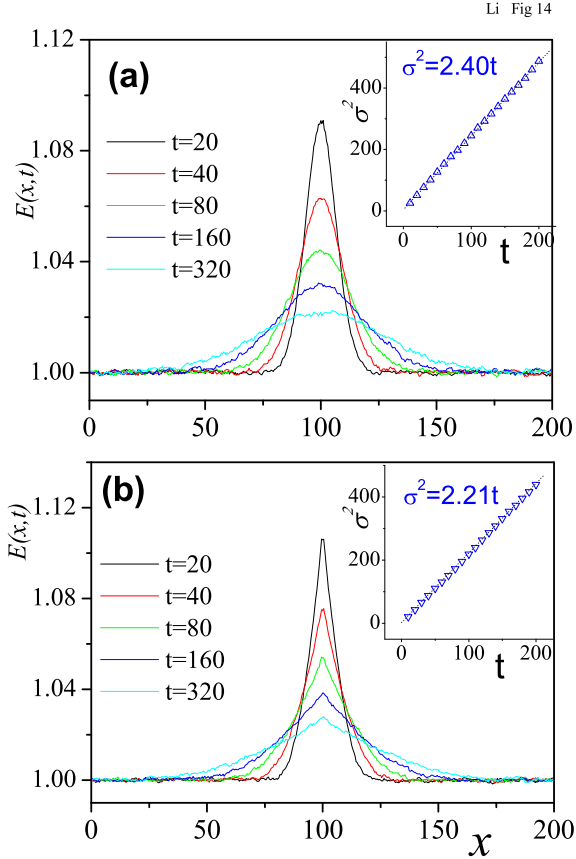}
\vspace{-.5cm}
\caption{Energy diffusion of the alternate mass hard-core potential model. (a) $M=1$, $m=(\sqrt{5}-1)/2, (b) m=M.$}
\label{fig:BBBdiffusion}
\end{figure}

We shall stress an important property of diffusion in this model. For any $m\ne M$ the system shows very nice normal Gaussian diffusion, see Fig.\ref{fig:BBBdiffusion}(a) for
$M=1, m=(\sqrt{5}-1)/2$. However for $m=M$ we found that even if $\sigma^2(t)=2Dt$, the probability distribution function (PDF) is not a Gaussian, instead, it can be described by the empirical formula:
$
\Delta E(x,t)=\Delta E(x_0,t)\exp \left(-0.826\left|x/\sigma(t)\right|^{1.69}\right),
$
with 
$
\sigma^2(t)=2.21t.
$
Except the two constants, this formula is same as that one for superdifussion process, Eq. (\ref{eq:SuperDfPDF}). Why these different processes have similar PDF is still an open question which deserves further investigations.

\section{Comparison between theory and numerical results}

\begin{figure}
\includegraphics[width=8.5cm]{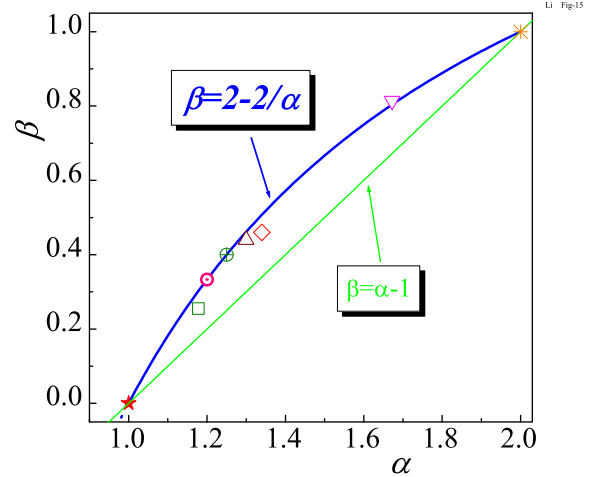}
\vspace{-.5cm}
\caption{\label{fig:alpha-beta}
The $\alpha-\beta$ plot. $\star$, normal diffusion (see the text for more explanations).
$\ast$, the ballistic transport.  $\bigtriangledown$, 1D Ehrenfest gas
channel\cite{LWH02};
$\square$, the rational triangle channel\cite{LCW03}; $\triangle$, the polygonal billiard channel
with $\phi_1=(\sqrt{5}-1)\pi/4$) and $\phi_2=\pi/3$\cite{Alonso02}; $\Diamond$, the triangle-square
channel gas. $\beta$ values are obtained from system size $L\in [192,384]$ for all channels except Ehrenfest channel\cite{LWH02}. $\odot$ the FPU lattice model at high temperature regime, and $\oplus$ the single walled nanotubes at room temperature.
For comparison, we also draw  Denisov \textsl{et al}'s result $\beta=\alpha-1$ (dotted line)\cite{Klafter}, and our results $\beta=2-2/\alpha$ (Eq.\ref{alpha-beta}) (solid curve).
}
\end{figure}

All numerical results are summarized and represented in Fig. \ref{fig:alpha-beta},
where we draw $\beta$ versus $\alpha$ for almost all known models, and compare it with 
Eq. (\ref{alpha-beta}). 
Here we give a detail discussion by classifying the model in four categories:
Ballistic, normal diffusion, superdiffusion, and suberdiffusion.

{\it A ballistic motion}, $\alpha=2$, leads to a divergent thermal
conductivity $\kappa \propto L$. The only existing analytical
result is heat conduction in a 1D harmonic lattice. It is known
that heat is transported by phonons in lattice model. Because
the absence of resistance and umklapp process, the phonons transport
ballistically in harmonic lattice model, thus $\alpha =2$. From
our formula (\ref{alpha-beta}), the thermal conductivity in the
1D harmonic lattice diverges as $L^{\beta}$ with $\beta=1$, this
is exactly what was shown by Lebowitz et al\cite{Lebowitz}
(``$\ast$" in Fig. \ref{fig:alpha-beta}).

{\it A normal diffusion}, $\alpha=1$, means that the thermal
conductivity is a size independent constant, $\beta=0$,
i.e. the heat conduction obeys the Fourier law. For example in the
1D Frenkel-Kontorova model, under certain condition (see Fig.\ref{fig:FKsigma2t}), the heat energy transports diffusively, and the thermal conductivity is finite
and independent of the system size $L$. Other 1D models
showing normal diffusion and normal thermal conduction are: the 1D
Lorentz gas channel\cite{Alonso99}, the 1D disordered Ehrenfest gas
channel\cite{LWH02}, the 1D irrational triangle
channel\cite{LCW03}, the alternate mass hard-core potential model\cite{LCWP04}, 
and some 1D polygonal billiard channels with irrationa $\phi_1$ and
certain rational triangle $\phi_2$ \cite{Alonso02}. 1D disordered lattice model under certain boundary conditions\cite{LZH01}. ``$\star$" in Fig. \ref{fig:alpha-beta}
represents all models with normal diffusion.

{\it A superdiffusion}, $1<\alpha<2$, implies an anomalous heat
conduction with a divergent thermal conductivity $L^{\beta}$. The
exponent $0<\beta<1$ differs from model to models. In billiard gas models, 
we have the Ehrenfest
gas channel\cite{LWH02}, the rational triangles model\cite{LCW03}, and the triangle-square channel. In lattice model, we have the FPU model and the nanotube model. As is shown in section IIIA and IVB for the FPU and the SWNT, respectively, the $\alpha$ and $\beta$ values agree with Eq.(\ref{alpha-beta}) perfectly. (See also Fig.\ref{fig:alpha-beta})

{\it A subdiffusion}. So far only one set data in this category is available, for which we have discussed already in Section VE.  More models worth studying in this case both for billiard channels and for lattice. In fact, to find a lattice with subdiffusion will be an interesting problem. If such kind of lattices exist, they must be a thermal insulator.

\subsection{Another scaling?}

Denisov et al\cite{Klafter} has also proposed a scaling relation between the anomalous diffusion and anomalous heat conduction for the so-called dynamic channel. According to their scaling,  if the particle moves in a superdiffusive way, then the thermal conductivity of the channel diverges as the system size $L$ as, $L^{\beta}$, with $\beta=\alpha -1$. They state that the billiard gas channels\cite{Alonso99,LWH02,Alonso02,LCW03} belong to the class of dynamical heat channels. Here we demonstrate with numerical evidence that dynamical heat channel might be useful for some unknown models, but not valid for the billiard gas model and lattice model. For comparison, we also draw $\beta=\alpha-1$ in Fig.\ref{fig:alpha-beta}. The deviation is clearly seen.

\begin{figure}
\includegraphics[width=8.5cm]{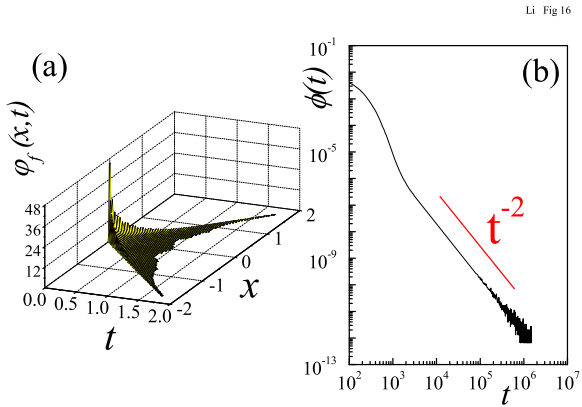}
\vspace{-.5cm}
\caption{\label{PDF}(a) The spatiotemporal PDF of the flight for the polygonal billiard channel\cite{Alonso02} with parameter $\phi_2=\pi/3$. 
Note that $\varphi_f(x,t)$ goes to zero for $t>1.8$,which is different from the assumption of the power law decay in $t$ as assumed by Denisov \textsl{et al.}\cite{Klafter}.
 (b) The PDF of the first passage time for the same billiard with cell umber $N=4$. It shows that $\phi(t)$ decays in a power law rather than in exponential way as assumed in Ref.\cite{Klafter}.
}
\end{figure}

In Fig \ref{PDF}(a), we show the spatiotemporal probability distribution function (PDF) of the flight (between two consecutive collisions with the channel boundaries) for the polygonal billiard channel with $\phi_2=\pi/3$\cite{Alonso02}. This figure shows that the situation in billiard channel is more complicated than that described by Levy walk  used in eq. (7) in Ref\cite{Klafter}. In the Levy walk, there is  no correlation between two consecutive walks, one can thus expect a probability of $2^{-n}$ for a particle to move in the same direction (left or right) in $n$ consecutive walks. However, for the polygonal billiard channel given here, we observed that this probability drops to zero as long as $n>6$. We also found that the probability for a particle to `hop' $n$ cells consecutively to left (or right) deviates remarkably from $2^{-n}$. This implies that the Levy walk model used in Ref.\cite{Klafter} oversimplifies the billiard gas channel.  

As a result, there exists a fundamental difference in PDF of the first passage time (FPT) for Levy walk model and  the billiard gas channels. The PDF of FPT $\phi(t)$ for the same billiard model is found to decay asymptotically in power law with slope of 2 (see Fig \ref{PDF} b) rather than in exponential way in Levy walk model. 

Based on the above two facts, it is not a surprise to find that there are large deviations of the numerical $\beta$ values from the formula $\beta=\alpha-1$ (dotted line), see Fig \ref{fig:alpha-beta}. 

In fact, the large deviation from the model used in Ref.\cite{Klafter} may result from inapproriate use of Levy walk.
Levy walk is characterized by two assumptive distribution functions which by specifying them appropriately, one can find a broad application spectrum for this statistical model. However, as is stressed  by Zaslavsky \cite{Zaslavsky02}, great care must be taken in specifying these assumptive distribution functions (such as $\psi_f(x,t)$ mentioned above) so that they do capture the characteristics of the dynamics. Failing to do so might be one of the reasons why $\beta=\alpha-1$ obtained in Ref.\cite{Klafter} can not be applied to the gas channels. Another reason is, in Levy walk it is assumed that between two successive steps there is no correlations; this is in clear contrast with the gas channels. This leads us to believe that by using the Levy walk in an appropriate way (which the main dynamic features are incorporated and the correlations are well dealt with) one can get deeper insight of the transport behaviors in the gas channels and lattices.

\section{Conclusions}
In summary, we have established a connection between anomalous
heat conduction and anomalous diffusion in 1D systems. The two exponents in anomalous difussion and anomalous thermal conductivity is connected via formula (\ref{alpha-beta}). This formula, includes all possible cases
observed in different classes of 1D models, ranging from
subdiffusion, normal diffusion, and superdiffusion to ballistic
transport. Several conclusions can be drawn: (1) A normal
diffusion leads to a normal heat conduction obeying the Fourier
law. (2) A ballistic transport leads to an anomalous heat
conduction with a divergent thermal conductivity  $\kappa\propto
L$. (3) A superdiffusion leads to an anomalous heat conduction
with a divergent thermal conductivity in thermodynamic limit. (4)
More importantly, our result predicts that a subdiffusion system
will be a thermal insulator. Existing numerical data from different low dimensional systems such as one dimensional lattice, small radius carbon nanotube and billiard gas channels support our results.

\bigskip
We would like to thank D K Campbell, P Rosenau and G Zaslavsky for the invitation to contribute this article to the CHAOS focus issue commenorating the 50'th anniversary of Fermi-Pasta-Ulam's model. Helpful comments and suggestions from G. Zaslavsky are very much appreciated. BL thanks G. Casati, T. Prosen, J.-S Wang and H Zhao for collaboration in some of the projects in past years during different periods. 
BL is supported in part by Academic Research Fund of NUS and the Temasek Young Investigator Award of DSTA Singapore under Project Agreement POD0410553. JW and LW are supported by DSTA
Singapore under Project Agreement POD0001821. GZ is supported by Singapore Millennium Foundation.

\end{document}